\documentstyle[11pt,nyas,epsfig]{article}

\newcommand{\beq}{\begin{equation}}
\newcommand{\eeq}{\end{equation}}
\newcommand{\beqa}{\begin{eqnarray}}
\newcommand{\eeqa}{\end{eqnarray}}

\newcommand{\lexp}{\mathop{\langle}}
\newcommand{\rexp}{\mathop{\rangle}}
\newcommand{\rexpc}{\mathop{\rangle_c}}

\def\d{\delta}

\def\dD{\delta_{\rm D}}

\font\BF=cmmib10
\def\k{{\hbox{\BF k}}}
\def\q{{\hbox{\BF q}}}

\def\v{{\hbox{\BF v}}}

\def\w{{\hbox{\BF w}}}

\def\fun#1#2{\lower3.6pt\vbox{\baselineskip0pt\lineskip.9pt
        \ialign{$\mathsurround=0pt#1\hfill##\hfil$\crcr#2\crcr\sim\crcr}}}

\begin{document}

\title{A New Angle on Gravitational Clustering}

\author{Rom\'an Scoccimarro}
\affil{Institute for Advanced Study, School of Natural Sciences, \\
Einstein Drive, Princeton, NJ 08540}

%
%
\begin{abstract}
We describe a new approach to gravitational instability in large-scale
structure, where the equations of motion are written and solved as in
field theory in terms of Feynman diagrams. The basic objects of
interest are the propagator (which propagates solutions forward in
time), the vertex (which describes non-linear interactions between
waves) and a source with prescribed statistics which describes the
effect of initial conditions. We show that loop corrections
renormalize these quantities, and discuss applications of this
formalism to a better understanding of gravitational instability and
to improving non-linear perturbation theory in the transition to the
non-linear regime. We also consider the role of vorticity creation due
to shell-crossing and show using N-body simulations that at small
(virialized) scales the velocity field reaches equipartition, i.e. the
vorticity power spectrum is about twice the divergence power spectrum.
\end{abstract}
%
%

%
%
\section{Standard Formulation of Gravitational Instability} 
\label{sec:STD}
%
%

Assuming the initial velocity field is irrotational, gravitational
instability can be described completely in terms of the density field
and the velocity divergence, $ \theta \equiv \nabla \cdot {\bf v }$.
Defining the conformal time $\tau=\int dt/a$ and the conformal
expansion rate $ {\cal H}\equiv {d\ln a /{d\tau}}$, the equations of
motion in Fourier space become

\beqa
  {\partial \tilde{\delta}(\k) \over {\partial \tau}} +
	\tilde{\theta}(\k) &=& - \int d^3 k_1 d^3 k_2 
	[\dD] \alpha(\k, \k_1) \tilde{\theta}(\k_1) 
	\tilde{\delta}(\k_2)   \label{ddtdelta}, \\
  {\partial \tilde{\theta}({\k}) \over{\partial \tau}} +
	{\cal H}\tilde{\theta}(\k) + {3\over 2}  
	\Omega {\cal H}^2\tilde{\delta}(\k) &=& 
   - \int d^3k_1 d^3k_2 [\dD] \beta(\k_1, \k_2)
	\tilde{\theta}(\k_1) \tilde{\theta}(\k_2)
	\label{ddttheta},
\eeqa

\noindent where $[\dD]=\dD(\k - \k_{12})$, $\k$ is a comoving wave
number, and 

\beq \alpha(\k, \k_1) \equiv {\k \cdot \k_1 \over{k_1^2}},
\qquad \beta(\k_1, \k_2) \equiv {k^2 (\k_1 \cdot \k_2 )\over{2 k_1^2
k_2^2}} \label{albe}.  
\eeq 

\noindent Equations (\ref{ddtdelta}) and (\ref{ddttheta}) are valid in
an arbitrary homogeneous and isotropic universe, which evolves
according to the Friedmann equations:

\beqa
  &&{\partial {\cal H}(\tau) \over {\partial \tau}} = - \frac{\Omega}{2}
	{\cal H}^2(\tau) + \frac{\Lambda}{3}a^2(\tau) \label{hdot}, \\
  &&(\Omega-1) {\cal H}^2(\tau) = k - \frac{\Lambda}{3}a^2(\tau)
  \label{keq},
\eeqa

\noindent where $\Lambda$ is the cosmological constant, the spatial
curvature constant $k=-1,0,1$ for $\Omega_{\rm tot}<1$, $\Omega_{\rm
tot}=1$, and $\Omega_{\rm tot}>1$, respectively, and $\Omega_{\rm
tot}\equiv\Omega+\Omega_\Lambda$, with $\Omega_\Lambda \equiv \Lambda
a^2/(3 {\cal H}^2)$.  For $\Omega=1$, the perturbative growing mode
solutions are given by
\beqa
	\tilde{\delta}(\k) &=& \sum_{n=1}^{\infty} a^n(\tau)
	\delta_n(\k)
	\label{ptansatz}, \\
	 \tilde{\theta}({\bf k}) &=&
	{\cal H}(\tau) \sum_{n=1}^{\infty} a^n(\tau) \theta_n({\bf k})
	\label{ptansatz2}.
\eeqa

\noindent Modelling the matter as pressureless non-relativistic `dust',
an appropriate description for cold dark matter before shell crossing,
the fluid equations of motion determine $\delta_n(\k)$ and $\theta_n(\k)$
in terms of the linear fluctuations,

\beqa
\delta_n(\k) &=& \int d^3q_1 \ldots \int d^3q_n [\dD]_n
 \ F_n^{(s)}(\q_1,  \ldots  ,\q_n)
\ \delta_1(\q_1) \ldots \delta_1(\q_n)
\label{ec:deltan}, \\
\theta_n({\k}) &=& - \int d^3q_1 \ldots \int d^3q_n [\dD]_n
G_n^{(s)}({\q}_1,  \ldots  ,{\q}_n)
\delta_1({\q}_1) \ldots \delta_1({\q}_n)
\label{ec:thetan},
\eeqa

\noindent where $[\dD]_n \equiv \dD(\k-\q_1 - \ldots - \q_n)$.  The
functions $F_n^{(s)}$ and $G_n^{(s)}$ are constructed from the mode
coupling functions $\alpha({\k}, {\k}_1)$ and $\beta({\k}_1, {\k}_2)$
by a recursive procedure (Goroff et al. 1986),

\beqa
F_n(\q_1,  \ldots , \q_n) &=& \sum_{m=1}^{n-1}
  { G_m(\q_1,  \ldots , \q_m) \over{(2n+3)(n-1)}} \Bigl[ (2n+1) 
  \alpha(\k,\k_1)  F_{n-m}(\q_{m+1}, \ldots , \q_n) \nonumber \\
&& \qquad + 2 \beta(\k_1, \k_2) 
  G_{n-m}(\q_{m+1}, \ldots , \q_n) \Bigr] \label{ec:Fn}, \\
G_n(\q_1,  \ldots , \q_n) &=& \sum_{m=1}^{n-1}
 { G_m(\q_1, \ldots , \q_m) \over{(2n+3)(n-1)}}
 \Bigl[3 \alpha(\k,\k_1)  F_{n-m}(\q_{m+1}, \ldots , \q_n) \nonumber \\
&& \qquad + 2n \beta(\k_1, \k_2)  
  G_{n-m}(\q_{m+1}, \ldots ,\q_n) \Bigr] \label{ec:Gn}
\eeqa

\noindent (where $ \k_1 \equiv \q_1 + \ldots + \q_m$, $\k_2 \equiv
\q_{m+1} + \ldots + \q_n$, $\k \equiv \k_1 +\k_2 $, and $F_1 = G_1
\equiv 1$). In Eqs.(\ref{ec:deltan}-\ref{ec:thetan}), $F_n^{(s)}$ and
$G_n^{(s)}$  are the symmetrized version of $F_n$ and $G_n$,
respectively. 

From these perturbative solutions a number of important results have
been derived in the literature, most of them regarding the tree-level
(leading-order) behavior of correlation functions, e.g. Fry (1984),
Goroff et al. (1986), Bernardeau (1992,1994). Loop calculations have
been attempted only in some particular cases (Scoccimarro \& Frieman
1996; Scoccimarro 1997, Scoccimarro et al. 1998); although in the
spherical collapse approximation a number of useful results have been
obtained (Fosalba \& Gazta\~naga 1998; Gazta\~naga \& Fosalba 1998).

%
%
\section{Field Theory Approach} 
\subsection{Integral Form of the Equations of Motion} 
\label{sec:IF}
%
%
The equations of motion can be rewritten in a more symmetric form by
defining a two-component ``vector'' $\Psi_a(\k,z)$, where $a=1,2$,
$z\equiv \ln a$, and: 

\beq
\Psi_a(\k,z) \equiv \Bigg( \delta(\k,z),\ -\theta(\k,z)/{\cal H} \Bigg),
\eeq
which for $\Omega=1$ leads to the following equations of motion (we
henceforth use 
the convention that repeated Fourier arguments are integrated over)

\beq
\partial_z \Psi_a(\k,z) + \Omega_{ab} \Psi_b(\k,z) =
 \gamma_{abc}(\k,\k_1,\k_2) \ \Psi_b(\k_1,z) \ \Psi_c(\k_2,z),
\label{eom}
\eeq
where 
\beq
\Omega_{ab} \equiv \Bigg[ 
\begin{array}{cc}
0 & -1 \\ -3/2 & 1/2 
\end{array}        \Bigg],
\eeq
and $\gamma_{abc}$ is a matrix given by

\beqa
\gamma_{121}(\k,\k_1,\k_2)&=&\dD(\k-\k_{12}) \ \alpha(\k,\k_1),
\label{ga1} \\
\gamma_{222}(\k,\k_1,\k_2)&=&\dD(\k-\k_{12}) \ \beta(\k_1,\k_2),
\label{ga2}
\eeqa

\noindent and zero otherwise. The somewhat complicated expressions for
the PT kernels recursion relations in the previous section can be
easily derived in this formalism. The perturbative solutions read

\beq
\Psi_a(\k,z) = \sum_{n=1}^\infty {\rm e}^{n z} \ \psi_a^{(n)}(\k)
\label{pta},
\eeq
which leads to

\beq
(n\d_{ab}+\Omega_{ab})\ \psi_b^{(n)}(\k) = 
\gamma_{abc}(\k,\k_1,\k_2) 
\sum_{m=1}^{n-1} \psi_b^{(n-m)}(\k_1) \ \psi_c^{(m)}(\k_2).
\eeq

Now, let $\sigma_{ab}^{-1}(n) \equiv n\d_{ab}+\Omega_{ab}$, then we have:

\beq
\psi_a^{(n)}(\k) = \sigma_{ab}(n) \ 
\gamma_{bcd}(\k,\k_1,\k_2) 
\sum_{m=1}^{n-1} \psi_c^{(n-m)}(\k_1) \ \psi_d^{(m)}(\k_2),
\label{GGRWm}
\eeq

\noindent where 

\beq \sigma_{ab}(n) = \frac{1}{(2n+3)(n-1)} \Bigg[ \begin{array}{cc}
2n+1 & 2 \\ 3 & 2n \end{array} \Bigg].  \eeq 

\noindent Equation (\ref{GGRWm}) is the equivalent to the Goroff et
al. (1986) recursion relations, Eqs.~(\ref{ec:Fn}-\ref{ec:Gn}), for
the $n^{\rm th}$ order Fourier amplitude solutions
$\psi_a^{(n)}(\k)$. It turns out to be convenient to write down the
equation of motion Eq.~(\ref{eom}) in integral form. Laplace
transformation in the variable $z$ leads to:

\beq
\sigma_{ab}^{-1}(\omega)\  \Psi_b(\k,\omega) = \phi_a(\k) +
\gamma_{abc}(\k,\k_1,\k_2) \oint \frac{d\omega_1}{2\pi i} \ 
\Psi_b(\k_1,\omega_1) \Psi_c(\k_2,\omega-\omega_1),
\label{eom2}
\eeq

\noindent where $\phi_a(\k)$ denote the initial conditions, that is
$\Psi_a(\k,z=0) \equiv \phi_a(\k)$. Multiplying by the matrix
$\sigma_{ab}$, and performing the inversion of the Laplace transform
we finally get

\beq
\Psi_a(\k,z) = g_{ab}(z) \ \phi_b(\k) + \int_0^z  dz' \ g_{ab}(z-z')
\ \gamma_{bcd}(\k,\k_1,\k_2)\ \Psi_c(\k_1,z') \Psi_d(\k_2,z'),
\label{eomi}
\eeq

\noindent where the {\em linear propagator} $g_{ab}(z)$ is defined as
($c>1$ to pick out the standard retarded propagator, Scoccimarro 1998) 

\beq
g_{ab}(z) = \oint_{c-i\infty}^{c+i\infty} \frac{d\omega}{2\pi i}
\sigma_{ab}(\omega) \ {\rm e}^{\omega z} = \frac{{\rm e}^z}{5}
\Bigg[ \begin{array}{rr} 3 & 2 \\ 3 & 2 \end{array} \Bigg] -
\frac{{\rm e}^{-3z/2}}{5}
\Bigg[ \begin{array}{rr} -2 & 2 \\ 3 & -3 \end{array} \Bigg],
\label{prop}
\eeq 

\noindent for $z\geq 0$, whereas $g_{ab}(z) =0$ for $z<0$ due to
causality, $g_{ab}(z) \rightarrow \d_{ab}$ as $z\rightarrow 0^{+}$.
The first term in Eq.~(\ref{prop}) represents the propagation of
linear growing mode solutions, where the second corresponds to the
decaying modes propagation. If we assume that the initial conditions are
set in the growing mode, then $\phi_a(\k) = \d_{1}(\k) (1,1)$
and the second term in Eq.~(\ref{prop}) does not contribute
upon contraction with $\phi_b(\k)$. Consistently with this, we can 
neglect subdominant time dependences in the non-linear term in
Eq.~(\ref{eomi}), which amounts to setting the initial conditions at
$z=-\infty$. Then, the equations of motion in integral form reduce to:

\beq
\Psi_a(\k,z) = 
{\rm e}^z \phi_a(\k) + \int_{-\infty}^z  dz' \ g_{ab}(z-z')
\ \gamma_{bcd}(\k,\k_1,\k_2)\ \Psi_c(\k_1,z') \Psi_d(\k_2,z').
\label{eomin}
\eeq

\noindent As it stands, this integral equation can be thought as an
equation for $\Psi_a(\k,z)$ in the presence of an ``external source''
$\phi_a(\k)$ with prescribed statistics. In particular, if we assume that
the initial conditions are Gaussian; then the statistical properties
of $\phi_a(\k)$ are completely characterized by its two-point
correlator 

\beq
\lexp \phi_a(\k) \ \phi_b(\k') \rexp = \dD(\k+\k') \
u_{ab} P(k),
\label{2pt}
\eeq

\noindent where $P(k)$ denotes the initial power spectrum of density
fluctuations and $u_{ab}=1$ for growing-mode initial conditions. From
Eq.~(\ref{eomin}), it is easy to verify that the ansatz in
Eq.~(\ref{pta}) leads to the recursion relations in Eq.~(\ref{GGRWm}).

Equation~(\ref{eomi}) has a simple interpretation. The first term
corresponds to the linear propagation from the initial conditions,
whereas the second term contains information on non-linear
interactions (mode-mode coupling). This corresponds to all the
interactions between waves that happen at time $z'$ (with $0 \leq z'
\leq z$) characterized by $\gamma_{bcd}$ and then propagated forward
in time from $z'$ to $z$ by the propagator $g_{ab}(z-z')$. We can also
write down the general solution as a perturbation series,

\beq 
\psi_a^{(n)}(\k,z)= \int \dD(\k-\k_{1\ldots n}) 
{\cal F}_a^{(n)}(\k_1,\ldots,\k_n;z) \d_1(\k_1) \ldots \d_1(\k_n), 
\eeq

\noindent where the kernels satisfy the usual recursion relations

\beq
{\cal F}_a^{(n)}(z) = \sum_{m=1}^n \int_0^z ds
g_{ab}(z-s) \gamma_{bcd}(\k,\k_1,\k_2) {\cal
F}_c^{(m)}(s) {\cal
F}_d^{(n-m)}(s).
\eeq

Interactions modify the linear propagator, leading to {\em propagator
renormalization}, so that the non-linear propagator defined by

\beq
G_{ab}(k,z)\ \dD(\k-\k') \equiv \left\langle 
\frac{\d \Psi_a(\k,z)}{\d \phi_b(\k')}
\right\rangle_c,
\eeq

\noindent reads
\beq
G_{ab}(k,z) = g_{ab}(z) + \sum_{n=1}^{\infty}  A_n(z)
\int d^3 q_1 P_1 \ldots d^3 q_n P_n \ \frac{\partial \overline{{\cal
F}}_a^{(2n+1)}}{\partial u_b} ,
\eeq

\noindent where $\overline{{\cal F}}_a^{(2n+1)}={\cal F}_a^{(2n+1)}
(\k,\q_1,-\q_1,\ldots,\q_n,-\q_n)$, $A_n(z) \equiv (2n-1)!!\exp
[(2n+1)z]$, and we defined $\phi_b \equiv (u_1,u_2)
\d_1(\k)$. Similarly the vertex is renormalized by non-linear
interactions as well,

\beq
\Gamma_{abc}(\k_1,\k_2,z)\ \dD(\k-\k_{12}) \equiv 
G_{bd}^{-1} G_{ce}^{-1} \left\langle \frac{\d^2
\Psi_a(\k,z)}{\d \phi_d(\k_1) \d \phi_e(\k_2)} \right\rangle_c ,
\eeq

\noindent and thus $\Gamma_{abc}= \gamma_{abc} + {\rm corrections}$.

\begin{figure}[t!]
\centering
\centerline{\epsfxsize=12. truecm \epsfysize=2. truecm 
\epsfbox{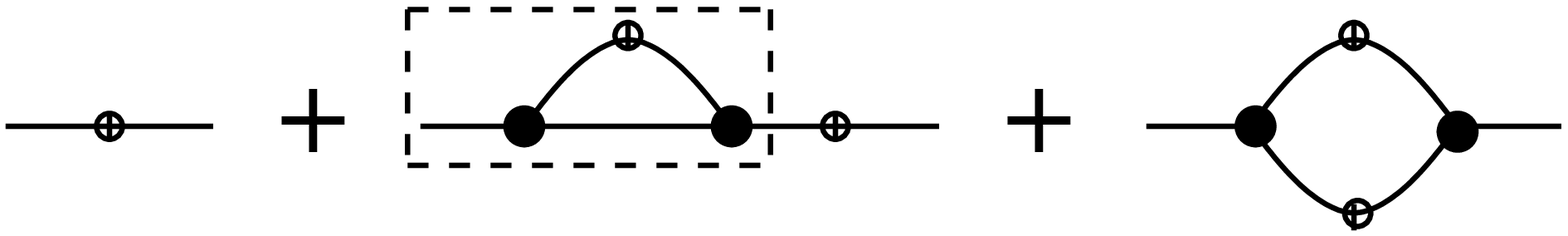}}
\caption{Power spectrum diagrams up to one-loop. The first term
denotes the linear contribution, the two remaining terms denote the
one-loop correction. The factor enclosed by dashed lines denotes
propagator renormalization. }
\label{fig_pk}
\end{figure}

\begin{figure}[t!]
\centering
\centerline{\epsfxsize=12. truecm \epsfysize=7. truecm 
\epsfbox{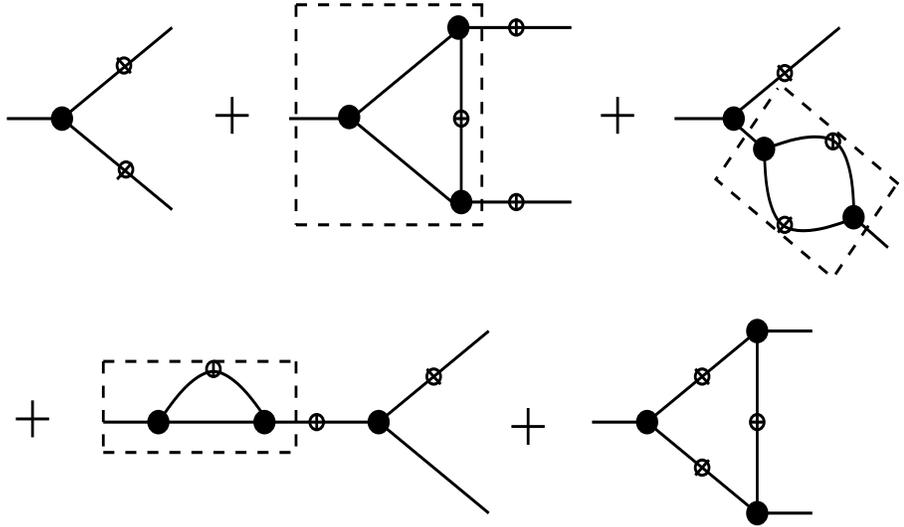}}
\caption{Bispectrum diagrams up to one-loop. The first terms denotes
the tree contribution, the four remaining terms the one-loop
correction. The first factor enclosed by dashed lines denotes vertex
renormalization, the second corresponds to the irreducible one-loop
power spectrum, the third denotes propagator renormalization. The last
term gives the irreducible one-loop bispectrum.}
\label{fig_bisp}
\end{figure}

The calculation of correlation functions can be written down in terms
of Feynman diagrams (Figs.~\ref{fig_pk}-\ref{fig_bisp}). We assign a
solid line to each propagator, Eq.~(\ref{prop}), a crossed circle
represents the two-point correlator in the initial conditions,
Eq.~(\ref{2pt}), and a solid circle represent the vertex,
Eqs.~(\ref{ga1}-\ref{ga2}). In this representation, loop corrections
can be divided into two general classes, those which renormalize the
propagator, and those which renormalize the vertex. For example, in
the calculation of the one-loop power spectrum there are two
contributions, $P^{(1)}= \lexp \d_1 \d_3 + \d_2 \d_2\rexpc$ (where
$\d_i$ denotes the i$^{\rm th}$ order solution in PT); the $\lexp \d_1
\d_3 \rexpc$ term corresponds to renormalizing the propagator (first
one-loop term in Fig.~\ref{fig_pk}), whereas the $\lexp \d_2 \d_2
\rexpc$ term denotes the irreducible one-loop power spectrum (second
one-loop term in Fig.~\ref{fig_pk}). By irreducible, we mean that this
contribution cannot be separated into two connected diagrams by
cutting one internal propagator line, unlike the $\lexp \d_1 \d_3
\rexpc$ contribution.

For the bispectrum, the 4 one-loop terms can be divided in a similar
fashion. The $\lexp \d_4 \d_1^2 \rexpc$ corresponds to vertex
renormalization (first one-loop term in Fig.~\ref{fig_bisp}), the
$\lexp \d_3 \d_2 \d_1 \rexpc$ correspond to power spectrum (second
one-loop term in Fig.~\ref{fig_bisp}) and propagator renormalization
(third one-loop term in Fig.~\ref{fig_bisp}), and the $\lexp \d_2^3
\rexpc$ gives the irreducible one-loop bispectrum (last term in
Fig.~\ref{fig_bisp}). This formalism can also be extended to include
non-Gaussian initial conditions, see Scoccimarro (1998) for a general
discussion and the specific example of Zel'dovich approximation
initial conditions, relevant to transients in N-body simulations.

%
%
\section{One-Loop Propagator and The Non-Linear Power Spectrum} 
\label{sec:res}
%
%

As an example, we calculate the one-loop propagator, $G_{ab}^{(1)}$

\beq
G_{ab}^{(1)}(k,z) = g_{ab}(z) + \exp(3z) \int d^3q P(q) \frac{\partial
{\cal F}^{(3)}_a}{\partial u_b}(\k,\q,-\q;z),
\eeq

If we take the $k\rightarrow 0$ limit, we find that (keeping only the
fastest growing term)

\beq
G_{ab}^{(1)}(k,z) = g_{ab}(z) - \sigma_v^2 \exp(3z) \Bigg[
\begin{array}{cc} 9/50 & 61/1050 \\ 3/25 & 61/1575 \end{array}
\Bigg] ,
\eeq

\noindent where $\sigma_v^2 \equiv \int P(q) d^3q/q^2$. Since the
correction is negative, this tends to make the non-linear growth
smaller than in linear theory, particularly for linearly decaying
modes, which decay faster than in the linear case. The correction to
$g_{ab}$ can be rewritten in terms of a correction to $\Omega_{ab}$,
using that

\beq
(\partial_z G_{ab})\ G_{bc}^{-1} = - \Omega_{ac},
\eeq

\noindent so that

\beq
\d\Omega_{ab} \approx k^2 \sigma_v^2\ \exp(7z/2)\   \Bigg[
\begin{array}{cc} -28/375 & 28/375 \\ -4/75 & 4/75 \end{array}
\Bigg].
\label{domega}
\eeq

In order to see the role of decaying modes in the standard solutions
of non-linear PT, let's consider the usual second-order PT kernel
(e.g. relevant for the calculation of the skewness and bispectrum). In
our notation, the second-order kernel can be written as

\beq
{\cal F}_1^{(2)}(\k_1,\k_2) = \int_0^z ds\ g_{1b}(z-s)\ \gamma_{bcd}\ \exp(2s)
\ (1,1)_c\ (1,1)_d,
\label{F2}
\eeq

\noindent where we assumed linear growing mode initial
conditions. Since $g_{1b}\gamma_{bcd}=
g_{11}\gamma_{121}+g_{12}\gamma_{222}$, and using Eq.~(\ref{prop}) we
have for the fastest growing contribution to 2nd-order PT 

\beqa
{\cal F}_1^{(2)}(\k_1,\k_2) &=& \exp(2z) \Bigg[
\Big(\frac{3}{5}+\frac{4}{35}\Big) \alpha(\k,\k_1) + 
\Big(\frac{2}{5}-\frac{4}{35}\Big) \beta(\k_1,\k_2) \Bigg] \nonumber
\\
&=& \exp(2z) \Big[
\frac{5}{7}  \alpha(\k,\k_1) + 
\frac{2}{7} \beta(\k_1,\k_2) \Big]. 
\eeqa

\noindent It is crucial to note here that the 4/35 contributions come
from {\em linearly decaying} modes; that is, after scattering, waves
are not in linearly growing modes anymore, and this type of amplitude
propagated into the present time contributes 4/35 to the amplitudes of
second-order PT kernels. That means that if we are using the kernels
to calculate one-loop corrections, which are important at
intermediate $k$'s, one could use the approximation in
Eq.~(\ref{domega}) to improve the propagator in Eq.~(\ref{F2}); in
this case this corresponds to supressing the linearly decaying mode
contribution to the propagator. As a result, the 2nd-order PT kernel
at intermediate scales would look more like

\beq
{\cal F}_1^{(2)}(\k_1,\k_2) = \exp(2z) \Big[
\frac{3}{5}  \alpha(\k,\k_1) + 
\frac{2}{5} \beta(\k_1,\k_2) \Big]. 
\eeq

\noindent This means that a simple way of improving one-loop
corrections in PT, is to use the kernels obtained by ignoring the
linearly decaying modes contributions. This has the effect of
incorporating higher-order loop corrections (those corresponding to
propagator renormalization, although only approximately since we don't
use the full one-loop propagator) in the usual formulation of PT.

\begin{figure}[t!]
\centering
\centerline{\epsfxsize=12. truecm \epsfysize=10. truecm 
\epsfbox{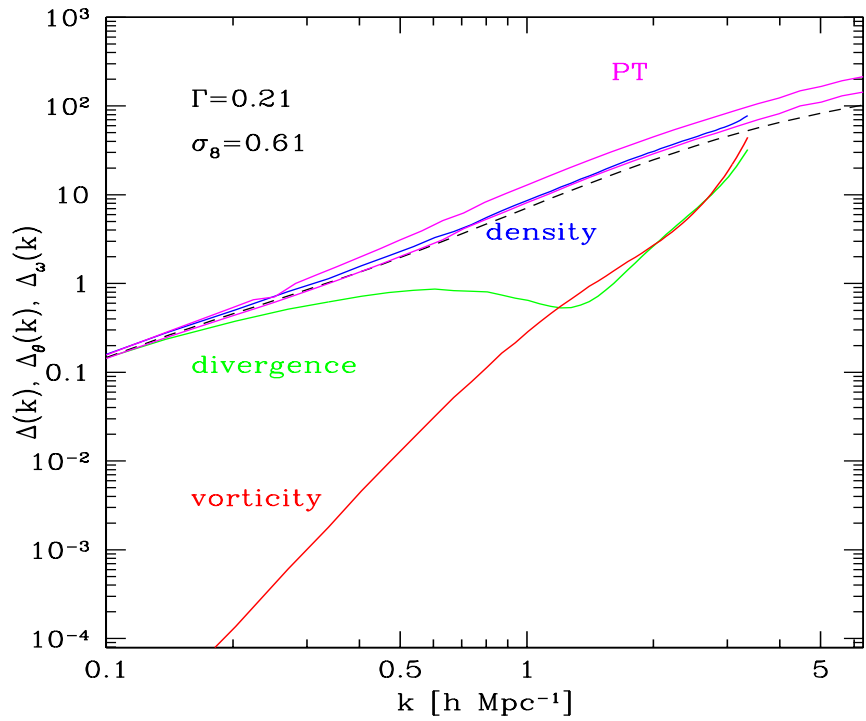}}
\caption{The power spectrum of the density, velocity divergence and
vorticity as a function of scale. The two solid lines roughly parallel
at high-k are the standard one-loop PT density power spectrum
calculation (top) and the new one-loop approach (bottom). The dashed
line denotes the prediction of the fitting formula and the solid line
close to it the actual measurement in the N-body simulation. The two
other solid lines denote the power spectrum of the velocity divergence
and vorticity, as labeled.}
\label{fig_pktcdm}
\end{figure}

\begin{figure}[t!]
\centering
\centerline{\epsfxsize=12. truecm \epsfysize=10. truecm 
\epsfbox{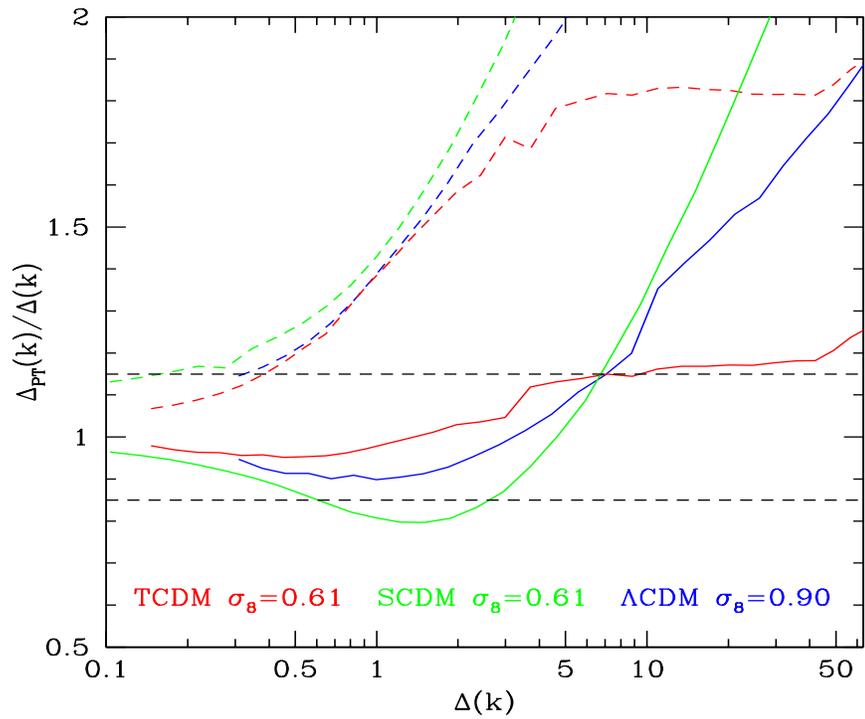}}
\caption{Ratio of predictions from one-loop PT (standard in dashed
lines, new approach in solid lines) to the fitting formula for the
non-linear power spectrum. The three different curves are for three
different models, as labeled.}
\label{fig_psratio}
\end{figure}

If we supress linearly decaying mode contributions for the third-order
kernel and use this to calculate one-loop corrections to the power
spectrum, we find the results in
Figs.\ref{fig_pktcdm}-\ref{fig_psratio}.  The solid lines in
Fig.~\ref{fig_pktcdm} show the standard (top) and ``improved''
(bottom) calculations of the non-linear power spectrum, whereas the
dashed line shows the fitting formula for the non-linear power
spectrum. The three-remaining solid lines (which extend up to $k
\approx 3$ h/Mpc) denote the measurement in N-body simulations of the
density power spectrum, the velocity divergence power spectrum and the
velocity vorticity power spectrum, as labeled. We see that the
vorticity power spectrum is certainly negligible at large scales, and
it does not become significant until scales of order $k \approx 2$
h/Mpc. At small scales, we find that the vorticity spectrum is roughly
twice that of the divergence, as expected if the velocity field has
equal power in all directions relative to $\k$.

Note that the ``improved'' calculation is somewhat smaller than the
standard one-loop calculation, as expected since the contribution from
propagating linearly decaying modes has been suppressed. Overall the
agreement with the N-body results is better. In Fig.~\ref{fig_psratio}
we show the ratio of our predictions for different models to the
non-linear fitting formula, the horizontal dashed lines show the
expected accuracy of the latter. We see that the ``improved''
calculations (solid) stay within the non-linear fitting formula
accuracy up to $k \approx 5$ h/Mpc, whereas the standard one-loop
calculation (dashed) overestimates the non-linear power spectrum at
scales smaller than the non-linear scale, $k \approx 0.3$
h/Mpc. The improvement is thus quite significant, although we have
included the effects of propagator renormalization in a crude way.

Unlike the velocity divergence, which can be calculated in one-loop PT
in analogous fashion to the density power spectrum, understanding the
vorticity power spectrum is considerably more complicated because
vorticity is generated by shell-crossing, an effect which is
negelected in the formulation of PT (see e.g. Pichon \& Bernardeau,
1999). However, we can understand approximately the scaling with
redshift and scale from simple considerations. After shell-crossing,
vorticity develops because what we see is the mass average of
different streams, each with its own (irrotational) velocity
field. Thus, vorticity can be thought as coming from the vorticity of
the mass-weighted velocity field, i.e.

\beq
\w \sim f_v(\tau)\ \nabla \times [(1+\d) \v],
\eeq

\noindent where $f_v(\tau)$ is the fraction of volume that undergoes
shell-crossing at time $\tau$. We can then write the vorticity power
spectrum ($\lexp \w(\k)\cdot \w(\k') \rexp \equiv P_w(k) \dD(\k+\k')$)

\beq
P_w(k) \sim f_v^2(\tau) \int \frac{|\k \times \q |}{q^4} \Big[
P_\theta(|\k-\q|) P_\d(q)- P_x(|\k-\q|) P_x(q) \Big] d^3q,
\eeq

\noindent where $P_\theta(k)$ is the velocity divergence power spectrum
and $P_x(k)$ is the power spectrum of the density-velocity divergence
cross-correlation. The simplest approximation would be to use linear
PT (although it is unlikely to be valid for each flow at the scales of
interest); however, since $P_\theta=P_\d=P_x$ in linear PT, this
contribution vanishes. Thus, the leading-order contribution to
$P_w(k)$ comes from one-loop PT,

\beq
P_w(k) \sim f_v^2(\tau) k^2 \int \frac{d^3q}{q^2} a^6 |k-q|^{2n+3} q^n
\sim a^6  f_v^2 k^{3n+6}.
\eeq

\noindent Thus, we expect a strong time and scale dependence for the
vorticity power spectrum. The latter is in reasonable agreeement with
Fig.~\ref{fig_pktcdm}, the time dependence is more difficult to test
due to the unknown dependence coming from $f_v^2(\tau)$.

%
%
\section{Conclusions} 
\label{sec:conc}
%
%

We described a new approach to gravitational instability in
large-scale structure, where the equations of motion are written and
solved as in field theory in terms of Feynman diagrams. The basic
objects of interest are the propagator (which propagates solutions
forward in time), the vertex (which describes non-linear interactions
between waves) and a source with prescribed statistics which describes
the effect of initial conditions. Loop corrections renormalize these
quantities, in particular, decaying modes are supressed in the
one-loop propagator compared to linear PT. We used this to construct
the PT kernels and calculate ``improved'' loop corrections, these
include effects beyond standard one-loop PT calculations, leading to
better agreement with N-body simulations for the evolution of the
power spectrum. We also consider the role of vorticity creation due to
shell-crossing and show using N-body simulations that at small
(virialized) scales the velocity field reaches equipartition, i.e. the
vorticity power spectrum is about twice the divergence power
spectrum. We also sketched a derivation of the time dependence and
scaling of the vorticity power spectrum.

Our calculations are only a first attempt to include the effects of
propagator renormalization, more work is needed to confirm that the
results presented here are indeed robust to a more careful
treatment. We have also neglected vertex renormalization. On the other
hand, it seems that this approach can lead to useful insights into the
nature of non-linear corrections and perhaps give us a more accurate
way to calculate clustering statistics in the transition to the
non-linear regime. Our results on the vorticity from numerical
simulations suggest that there is a significant range of scales until
the assumption of irrotational fluid breaks down. We hope that by
using these techniques we can finally answer the question: ``Why does
PT work so well?''

%
%
\section {Acknowledgements}
I thank Francis Bernardeau and Dmitry Pogosyan for useful discussions.

%
%

\clearpage

\end{document}